\documentclass[12pt]{article}
\begin{document}
\setlength{\unitlength}{0.2cm}
\date{}
\title{
High-precision determination of the critical exponent $\gamma$
for self-avoiding walks
}
\author{
  \\
  {\small Sergio Caracciolo}             \\[-0.2cm]
  {\small\it Scuola Normale Superiore and INFN -- Sezione di Pisa}  \\[-0.2cm]
  {\small\it I-56100 Pisa, ITALIA}          \\[-0.2cm]
  {\small Internet: {\tt Sergio.Caracciolo@sns.it}}     \\[-0.2cm]
  \\[-0.1cm]  \and
  {\small Maria Serena Causo}             \\[-0.2cm]
  {\small\it Dipartimento di Fisica and INFN -- Sezione di Lecce}  \\[-0.2cm]
  {\small\it Universit\`a degli Studi di Lecce}        \\[-0.2cm]
  {\small\it I-73100 Lecce, ITALIA}          \\[-0.2cm]
  {\small Internet: {\tt causo@le.infn.it}}     \\[-0.2cm]
  \\[-0.1cm]  \and
  {\small Andrea Pelissetto}                          \\[-0.2cm]
  {\small\it Dipartimento di Fisica and INFN -- Sezione di Pisa}    \\[-0.2cm]
  {\small\it Universit\`a degli Studi di Pisa}        \\[-0.2cm]
  {\small\it I-56100 Pisa, ITALIA}          \\[-0.2cm]
  {\small Internet: {\tt pelisset@ibmth.difi.unipi.it}}   \\[-0.2cm]
  {\protect\makebox[5in]{\quad}}  
  \\
}
\vspace{0.5cm}
\maketitle
\thispagestyle{empty}   
\vspace{0.2cm}
\begin{abstract}
We compute the exponent $\gamma$ for self-avoiding walks in three 
dimensions. We get $\gamma = 1.1575 \pm 0.0006$ in agreement with 
renormalization-group predictions. Earlier Monte Carlo 
and exact-enumeration determinations are now seen to be biased by 
corrections to scaling.
\end{abstract}
\clearpage
\newcommand{\be}{\begin{equation}}
\newcommand{\ee}{\end{equation}}
\newcommand{\bea}{\begin{eqnarray}}
\newcommand{\eea}{\end{eqnarray}}
\newcommand{\<}{\langle}
\renewcommand{\>}{\rangle}
\def\spose#1{\hbox to 0pt{#1\hss}}
\def\ltapprox{\mathrel{\spose{\lower 3pt\hbox{$\mathchar"218$}}
 \raise 2.0pt\hbox{$\mathchar"13C$}}}
\def\gtapprox{\mathrel{\spose{\lower 3pt\hbox{$\mathchar"218$}}
 \raise 2.0pt\hbox{$\mathchar"13E$}}}
\newcommand{\R}{\hbox{{\rm I}\kern-.2em\hbox{\rm R}}}
\newcommand{\reff}[1]{(\ref{#1})}
The self-avoiding walk (SAW) is a well-known lattice model of a polymer 
molecule in a good solvent;
moreover, because of its semplicity, it is an important test-case in the 
theory of critical phenomena. 
Much work has been done in computing critical exponents by a variety
of theoretical approaches -- Monte Carlo (MC),
exact enumeration/extrapolation (EE), and renormalization group (RG) --
with the aim of comparing these determinations with each other 
and with the experimental results. Small but persistent discrepancies 
have emerged between the predictions from different theoretical
approaches. For example, extensive studies have been done on the 
exponent $\nu$ which controls the critical behaviour of the 
length scales. Early MC simulations and EE 
studies \cite{Watts} in the '70s predicted $\nu=3/5$ in 
agreement with the Flory theory. When the length of the 
walks which were simulated was increased 
\cite{Rapaport,Sokal1,Eizenberg-Klafter}
the value of $\nu$ decreased to $\nu = 0.592 \pm 0.002$. This value 
was also supported by extended EE studies 
which provided an identical estimate \cite{Guttmann}. 
On the other hand, field-theoretic RG computations 
\cite{RG-francesi0,RG-francesi1,RG-francesi2,Muthu-Nickel,Nickel2}
persistently gave $\nu = 0.588 \pm 0.001$ or even slightly lower.
The discrepancy was clarified when the MC studies were extended to 
much longer walks: because of strong corrections
with non-analytic exponent $\Delta\approx 0.5$, the asymptotic regime is 
reached only for very long chains, and the results from shorter 
chains are systematically biased upward \cite{Majid-et-al,Zifferer,CFP,Sokal2}.
A simulation \cite{Sokal2} with walks of length up to 
$N=80000$, using a data analysis taking careful account of the 
corrections to scaling, gave $\nu=0.5877\pm 0.0006$ in good agreement
with the renormalization-group estimates. 
Also universality is well satisfied: 
a simulation \cite{CFP} in a slightly different geometry provided 
$\nu = 0.5867 \pm 0.0013$ (68\% confidence limit) while a
recent high-statistics simulation \cite{Pedersenetal}
for the off-lattice Kratky-Porod model with excluded volume 
gave $\nu = 0.5880 \pm 0.0018$.
One may ask if the same phenomenon occurs for the other critical 
exponents. For the exponent $\gamma$, there are indeed significant 
discrepancies in the existing theoretical predictions. MC and EE 
studies provide the estimates 
\be
\gamma\, =\, \cases{ 1.161\pm 0.001 & \hskip 1truecm 
                                    EE,      Ref. \protect\cite{Guttmann} \cr
                     1.1619\pm 0.0001 & \hskip 1truecm 
                                    EE,      Ref. \protect\cite{MacDonald2} \cr
                     1.1608 \pm 0.0003 & \hskip 1truecm 
                                    MC,      Ref. \protect\cite{Grassberger} 
}
\label{gammaestimates}
\ee
On the other hand the $\epsilon$-expansion predicts a lower value
\cite{RG-francesi1}:
$1.157 \ltapprox \gamma \ltapprox 1.160$. 
Indeed a Borel-type resummation gives $\gamma = 1.160 \pm 0.004$ and 
$\gamma = 1.157 \pm 0.003$ if one forces 
the expansion to reproduce the exact value in two dimensions
$\gamma = {43\over32}$.
One can also use the scaling relation $\gamma = (2 - \eta)\nu$ and the 
estimates for $\eta$ and $\nu$: the unconstrained 
$\epsilon$-expansion gives $\nu = 0.5885\pm 0.0025$ and 
$\eta = 0.031 \pm 0.003$ so that $\gamma = 1.159 \pm 0.005$;
using the exactly known values for $d=2$, $\nu={3\over4}$ and 
$\eta = {5\over24}$, one gets $\nu = 0.5880\pm 0.0015$ and 
$\eta = 0.0320 \pm 0.0025$ so that $\gamma = 1.1572 \pm 0.0035$.
More controversial is the status of the expansions at fixed spatial 
dimension $d=3$. In
\cite{RG-francesi1} $\gamma$ is estimated as $\gamma = 1.1615 \pm 0.0020$
while in \cite{Muthu-Nickel} the final estimate is 
$\gamma \approx 1.1613$. 
These estimates depend crucially on the 
critical value of the renormalized coupling constant $\overline{g}^*$:
in \cite{RG-francesi0,RG-francesi1} the estimate
$\overline{g}^* = 1.421 \pm 0.008$
is used , while \cite{Muthu-Nickel} uses
$\overline{g}^* = 1.422 \pm 0.008$.
However Nickel \cite{Nickel} has pointed out that the present estimates 
of $\overline{g}^*$ could be slightly higher than the correct value 
due to a possible non-analyticity of the $\beta$-function at 
$\overline{g}^*$ which is usually neglected in the standard analyses.
A reanalysis of the series \cite{Nickel2} indicates
that $\overline{g}^*$ could be as low as $1.39$ and predicts
$\gamma = 1.1569 + 0.10 (\overline{g}^* - 1.39) \pm 0.0004$. 
In this paper we present a high-precision MC 
study for the exponent $\gamma$ using walks up to length 
$N=40000$. Our results confirm the important role
played by corrections to scaling; our final estimate 
$\gamma = 1.1575 \pm 0.0006$ 
is significantly lower than previous MC and EE
results but it is in perfect agreement with the RG predictions. Moreover 
it clearly supports the analysis by Nickel 
\cite{Nickel,Nickel2} and the value $\overline{g}^* \approx 1.39$ for the 
renormalized coupling constant.
In presence of strong corrections to scaling, in order to get a reliable
estimate of the critical exponents, one needs to perform the simulation 
in the large-$N$ regime.  This is only possible if the 
algorithm at hand does not exhibit too strong a 
critical slowing-down. For the study
of $\gamma$ for SAWs on the lattice the best available algorithm is 
the {\em join-and-cut} algorithm \cite{join-and-cut}: 
in two dimensions the autocorrelation
time, expressed in CPU units, behaves approximately as
$N^{\approx 1.5}$ while in three dimensions the behaviour is expected to 
be $N^{\approx 1.2}$. The algorithm is thus nearly optimal. 
Another advantage of this algorithm is that it does not require 
the determination of the connectivity constant, at variance with 
more standard algorithms. 
The join-and-cut algorithm works in the unorthodox ensemble $T_{N_{tot}}$
consisting of all pairs of SAWs (each walk starts at the origin and 
ends anywhere) such that the {\em total} number of steps in the two
walks is some fixed number $N_{tot}$. Each pair in the ensemble is given
equal weight: therefore the two walks are not interacting except for the 
constraint on the sum of their lengths.
One sweep of the algorithm consists of two steps:
\begin{enumerate}
\item Starting from a pair of walks $(\omega_1,\omega_2)$, 
we update each of them independently using some 
ergodic fixed-length algorithm. We use the pivot 
algorithm \cite{Lal,MacDonald,Sokal1} which is the best available one for the
ensemble of fixed-length walks with free endpoints.
\item We perform a {\em join-and-cut} move: we concatenate the two walks
$\omega_1$ and $\omega_2$ forming a new (not necessarily 
self-avoiding) walk $\omega_{conc}$; then we cut $\omega_{conc}$ at a random
position creating two new walks ${\omega'}_1$ and ${\omega'}_2$. 
If ${\omega'}_1$ and ${\omega'}_2$ are self-avoiding we keep them;
otherwise the move is rejected and we stay with $\omega_1$ and $\omega_2$.
\end{enumerate}
More details on the dynamical critical behaviour and on the 
implementation of this algorithm can be found in \cite{join-and-cut}.
Let us now discuss how the critical exponent $\gamma$ can be estimated from 
the Monte Carlo data produced by the join-and-cut algorithm.
Let us start by noticing that the random variable $N_1$, the length of the 
first walk, has the distribution
\be
{\overline\pi}(N_1) \, =\, {c_{N_1} c_{N_{tot}-N_1}\over Z(N_{tot})}
\ee
for $1\le N_1\le N_{tot}-1$ where $Z(N_{tot})$ is the obvious 
normalization factor and $c_N$ is the number of walks going from the 
origin to any lattice point with $N$ steps.
For large $N$ we have
\be
c_N\approx a_0 \mu^N N^{\gamma-1}
\label{cN}
\ee
and thus the idea is to make inferences of $\gamma$ from the observed 
statistics of $N_1$. Of course the problem is that \reff{cN} is an 
asymptotic formula valid only in the large-$N$ regime. We will thus 
proceed in the following way: we will suppose that \reff{cN} is valid
for all $N\ge N_{min}$ for many increasing values of $N_{min}$ 
and correspondingly we will get estimates $\hat{\gamma}(N_{min})$; 
these quantities are effective exponents which depend on $N_{min}$ and 
which give correct estimates of $\gamma$ as $N_{min}$ and 
$N_{tot}$ go to infinity.
The determination of $\gamma$ from the data is obtained using the 
maximum-likelihood method. We will present here only the 
results: for a detailed discussion we refer the reader to 
\cite{join-and-cut}.
Given $N_{min}$ consider the function 
\be
\theta_{N_{min}} (N_1) =\, 
   \cases{1 & if $N_{min}\le N_1\le N_{tot}$ \cr
          0 & otherwise}
\ee
and let $X$ be the random variable 
\be
X = \log[ N_1(N_{tot}-N_1)] \;\; .
\ee
Then define
\be
X^{cens}(N_{min}) = {\< X\theta_{N_{min}}\> \over \< \theta_{N_{min}}\>}
\label{Xcens}
\ee
where the average $\<\ \cdot\ \>$
is taken in the ensemble $T_{N_{tot}}$ sampled 
by the join-and-cut algorithm. The quantity defined in \reff{Xcens}
is estimated in the usual way from the Monte Carlo data obtaining is this 
way $X^{cens}_{MC} (N_{min})$.
Then $\hat{\gamma}(N_{min})$ is computed by solving the equation
\be
X^{cens}_{MC} (N_{min}) \, =\, [X]_{th,\hat{\gamma}} (N_{min})
\ee
where, for every function of $N_1$, we define
\be
[f(N_1)]_{th,\gamma} (N_{min}) \equiv 
  {\sum_{N_1 = N_{min}}^{N_{tot}-N_{min}} 
      f(N_1) N_1^{\gamma-1} (N_{tot}-N_1)^{\gamma-1} \over 
   \sum_{N_1 = N_{min}}^{N_{tot}-N_{min}} 
       N_1^{\gamma-1} (N_{tot}-N_1)^{\gamma-1} }
\ee
The variance of $\hat{\gamma}(N_{min})$ is then given by
\be
{\rm Var}\ [\hat{\gamma}(N_{min})] = \
   {{\rm Var}\ (X^{cens}_{MC} (N_{min}))\over 
    ([X;X]_{th,\hat{\gamma}} (N_{min}))^2 }
\ee
where $[X;X] = [X^2] - [X]^2$.
We must finally compute ${\rm Var}\ (X^{cens}_{MC} (N_{min}))$.
As this quantity is defined as the 
ratio of two mean values (see formula \reff{Xcens}) one must take into account
the correlation between denominator and numerator. 
Here we have used the standard formula for the variance of a ratio
(valid in the large-sample limit)
\be
{\rm Var}\left({A\over B}\right)\, =\, 
   {\<A\>^2\over \<B\>^2} {\rm Var} 
  \left( {A\over \<A\>} - {B\over \<B\>} \right)
\label{varformula}
\ee
Finally let us mention how to combine data from runs at different values of 
$N_{tot}$. The approach we use consists in analyzing the data separately for
each $N_{tot}$ and then in constructing the
usual weighted average of the resulting 
estimates $\hat{\gamma}(N_{min})$ with weights inversely proportional to 
the estimated variances.
Let us now discuss our results. We have performed high-statistics runs
at $N_{tot}=200,2000,20000$ and $40000$. The number of iterations is 
reported in Table \ref{tabella_tau}. The total simulation took 16 months on 
an AlphaStation 600 Mod 5/266. In the same table
we report also, for two different values of $N_{min}$,
the autocorrelation times for the observable
\be
Y(N_{min}) = {X\theta_{N_{min}}\over \<X\theta_{N_{min}}\>} -
           {\theta_{N_{min}}\over \<\theta_{N_{min}}\>}
\ee
which according to \reff{varformula} controls the errors on $\gamma$.
We use here a standard autocorrelation analysis \cite{Sokal1} with 
a self-consistent window of $15\tau_{int,Y}$, supplemented by 
an {\em ad hoc} prescription to take into account tails in the 
autocorrelation functions \cite{Sokal2}. The contribution of the tail 
to $\tau_{int,Y}$ amounts to approximately 20\% for the two larger
values of $N_{tot}$, 2\% for $N_{tot}=2000$, while for 
$N_{tot}=200$ we have been unable to identify any tail. 
The results are consistent with the expectation of 
$\tau_{int,Y}\sim N^{\approx 0.3}$ so that, keeping into account that
the CPU-time per iteration increases as $\sim N^{\approx 0.9}$,
we find that the CPU-time per independent walk increases roughly as 
$N^{\approx 1.2}$.
Our raw data are reported in Table \ref{tabella_gamma1} and
Table \ref{tabella_gamma2} where we give 
for various $N_{min}$ the estimates of $X^{cens}_{MC} (N_{min})$.
Let us now look at the results. Let us consider first $N_{tot}=200$.
We see here that the estimates of $\gamma$ increase with $N_{min}$ and for 
$N_{min} = 50$ they give $\gamma \approx 1.1605$. This value is in 
agreement with Monte Carlo studies \cite{Grassberger} 
performed in the same range of values of $N$ and
exact-enumeration studies \cite{Guttmann,MacDonald2}. 
Consider now $N_{tot} = 2000$. One can see that the estimates of 
$\gamma$ are lower and indicate $1.1580 \ltapprox \gamma \ltapprox 1.1585$: 
clearly the estimate 
at $N_{tot}=200$ was biased upward by the corrections to scaling. 
Let us now consider the results of Table \ref{tabella_gamma2}
where we give the estimates of $\gamma$ coming from the weighted average
of the results with $N_{tot}=20000$ and $N_{tot}=40000$. The estimates 
are extremely flat and 
agree within error bars from $N_{min}=1$ to $N_{min}=8000$; they indicate
$1.1574\ltapprox \gamma \ltapprox 1.1578$. Again the new estimate is 
lower than the previous ones. At least at $N_{tot} = 2000$ there are 
still systematic deviations which are larger than the statistical error.
Of course the same could be true for our results at the larger value of 
$N_{tot}$. The most straightforward way to solve the question would be to
have data at a higher value of $N_{tot}$, say $N_{tot} = 4 \cdot 10^5$, 
with comparable statistics. However this is not possible with 
current computer resources. We have thus simply tried to estimate the 
systematic bias by comparing the results with different $N_{tot}$ 
assuming that the systematic error vanishes as $N_{tot}^{-1/2}$
(this is the behaviour one expects if the corrections to $c_N$, 
formula \reff{cN}, vanish as $N^{-1/2}$). In this way we get 
an estimate of 0.0002-0.0003 as the residual bias on the data at 
$N_{tot} = 40000$. We have thus taken as our final estimate 
the value at $N_{min} = 2000$:
\be
\gamma = 1.1575 \pm 0.0003 \pm 0.0003 
\ee
where the first error is the statistical one (68\% confidence limits)
while the second is a subjective estimate of the error due to the corrections 
to scaling.
Let us finally discuss the determinations of the 
connectivity constant $\mu$ which is usually derived together with the 
estimate of $\gamma$. Restricting ourselves to the case of the cubic
lattice --- the discussion is completely analogous in the other cases ---
EE and MC studies give the following 
estimates for $\beta_c \equiv 1/\mu$:
\be
\beta_c =\cases{0.213496\pm 0.000004 & \hskip 1truecm
                             EE,       Ref. \protect\cite{Guttmann} \cr
                0.213499\pm 0.000001 & \hskip 1truecm
                             EE,       Ref. \protect\cite{MacDonald2} \cr
                0.213492\pm 0.000001 & \hskip 1truecm
                             MC,       Ref. \protect\cite{Grassberger}
               }
\label{muestimates}
\ee
However these results have probably a large systematic error
as the corresponding estimates of $\gamma$ are quite far from the 
correct value. To give an estimate of this effect we have considered the series
\be
S(\delta) =\, \left[ \sum_{N=0}^\infty c_N \beta^N\right]^{1\over \delta}
\ee
where $c_N$ is the number of walks of length $N$. If $\delta = \gamma$,
$S(\delta)$ has a simple pole as $\beta = \beta_c$ and thus, neglecting
subleading corrections, $\beta_c$ can be determined by a simple 
Pad\`e analysis. Using the values of $c_N$ reported in
\cite{Guttmann,MacDonald2}, we find
\be
\beta_c=\cases{ 0.2134987 \pm 0.0000001 & for $\delta = 1.1619$ \cr
                0.2134811 \pm 0.0000031 & for $\delta = 1.1576$
     }
\label{ourmuestimates}
\ee
where the ``error" indicates the spread of the Pad\`e approximants 
we have considered.
Thus if use the value of $\gamma$ given by \cite{MacDonald2}, 
see \reff{gammaestimates},
we get a value for $\beta_c$ in agreement with their estimate
\reff{muestimates}. However if we use our estimate of $\gamma$, 
we get a $\beta_c$ which is off by five to twenty times the stated
error bars. Of course we cannot trust \reff{ourmuestimates} as a good
estimate of $\beta_c$ as the analysis still does not take into account
corrections to scaling; however this simple exercise  shows that $\beta_c$
is lower than expected. It should also be noticed that this trend 
is already visible in \reff{muestimates} where $\beta_c$ 
obtained by MC is lower that the EE estimates.
\medskip
\bigskip
We thank Alan Sokal and Ettore Vicari for useful discussions.
\begin{table}
\protect\footnotesize
\begin{center}
\begin{tabular}{|c|c|c|c|}
\hline
$ N_{tot} $ & $N_{iter}$ 
  & $ \tau_{int,Y(1)}$
  & $ \tau_{int,Y(1000)}$ \\
\hline\hline
200 & $5 \cdot 10^8$ & $1.47970 \pm 0.00055$ & - \\
2000 & $6.2 \cdot 10^8$ & $2.808 \pm 0.022$ & - \\
20000 & $10^8$ & $6.96 \pm 0.18$ & $9.35 \pm 0.21$ \\
40000 & $8.5 \cdot 10^8$ & $8.80 \pm 0.23$ & $10.78 \pm 0.24$ \\
\hline
\end{tabular}
\end{center}
\caption{Number of iterations and autocorrelation times for the various values 
of $N_{tot}$.} 
\label{tabella_tau}
\end{table}
\begin{table}
\protect\footnotesize
\begin{center}
\begin{tabular}{|c|c|c||c|c|c|}
\hline
   & \multicolumn{2}{|c||}{$N_{tot} = 200$} &  
   & \multicolumn{2}{|c|}{$N_{tot} = 2000$} \\ 
\hline
    $N_{min}$  & $X^{cens}_{MC}$   & $\gamma$  &
    $N_{min}$  & $X^{cens}_{MC}$   & $\gamma$  \\
\hline
1  &  $ 8.701168 \pm 0.000052 $  &  $ 1.15288 \pm 0.00011 $ & 
1  &  $ 13.298266 \pm 0.000068 $  &  $ 1.15782 \pm 0.00013 $ \\  
10 &  $ 8.842084 \pm 0.000036 $  &  $ 1.15808 \pm 0.00021 $ &
100 &  $ 13.452571 \pm 0.000046 $  &  $ 1.15802 \pm 0.00028 $ \\
20 &  $ 8.943002 \pm 0.000026 $  &  $ 1.15866 \pm 0.00034 $ &
200 &  $ 13.551931 \pm 0.000033 $  &  $ 1.15811 \pm 0.00045 $ \\
30 &  $ 9.017434 \pm 0.000019 $  &  $ 1.15875 \pm 0.00053 $ &
300 &  $ 13.625478 \pm 0.000025 $  &  $ 1.15838 \pm 0.00071 $ \\
40 &  $ 9.074675 \pm 0.000014 $  &  $ 1.15999 \pm 0.00084 $ &
400 &  $ 13.682081 \pm 0.000018 $  &  $ 1.1598\hphantom{0} \pm 0.0011\hphantom{0} $ \\
50 &  $ 9.119083 \pm 0.000010 $  &  $ 1.1605\hphantom{0} \pm 0.0014\hphantom{0} $ &
500 &  $ 13.725970 \pm 0.000013 $  &  $ 1.1584\hphantom{0} \pm 0.0019\hphantom{0} $ \\
\hline
\end{tabular}
\end{center}
\caption{Raw data and estimates of $\gamma$ for $N_{tot}=200$ and 
$N_{tot}=2000$.}
\label{tabella_gamma1}
\end{table}
\begin{table}
\protect\footnotesize
\begin{center}
\begin{tabular}{|c||c|c||c|}
\hline
   $N_{min}$  & $X^{cens}_{MC}(20000)$& $X^{cens}_{MC}(40000)$ & $\gamma$  \\
\hline
    1  & $17.90254  \pm  0.00026 $  & $19.28850  \pm  0.00010   $& $ 1.15763  \pm  0.00018 $\\
  200  & $17.94330  \pm  0.00023 $  & $19.310211  \pm  0.000095   $& $ 1.15762  \pm  0.00021 $\\
  400  & $17.97715  \pm  0.00021 $  & $19.329406  \pm  0.000090   $& $ 1.15758  \pm  0.00024 $\\
  600  & $18.00687  \pm  0.00020 $  & $19.346998  \pm  0.000086   $& $ 1.15765  \pm  0.00026 $\\
  800  & $18.03376  \pm  0.00019 $  & $19.363292  \pm  0.000083   $& $ 1.15760  \pm  0.00028 $\\
 1000  & $18.05830  \pm  0.00017 $  & $19.378624  \pm  0.000080   $& $ 1.15764  \pm  0.00030 $\\
 1200  & $18.08095  \pm  0.00016 $  & $19.393077  \pm  0.000077   $& $ 1.15754  \pm  0.00032 $\\
 1400  & $18.10198  \pm  0.00015 $  & $19.406824  \pm  0.000074   $& $ 1.15749  \pm  0.00034 $\\
 1600  & $18.12172  \pm  0.00014 $  & $19.419918  \pm  0.000071   $& $ 1.15744  \pm  0.00036 $\\
 1800  & $18.14022  \pm  0.00013 $  & $19.432477  \pm  0.000069   $& $ 1.15755  \pm  0.00038 $\\
 2000  & $18.15760  \pm  0.00012 $  & $19.444488  \pm  0.000067   $& $ 1.15749  \pm  0.00040 $\\
 2200  & $18.17394  \pm  0.00012 $  & $19.456041  \pm  0.000065   $& $ 1.15746  \pm  0.00042 $\\
 2400  & $18.18930  \pm  0.00011 $  & $19.467169  \pm  0.000063   $& $ 1.15739  \pm  0.00045 $\\
 2600  & $18.20388  \pm  0.00011 $  & $19.477901  \pm  0.000061   $& $ 1.15741  \pm  0.00047 $\\
 2800  & $18.21771  \pm  0.00010 $  & $19.488255  \pm  0.000059   $& $ 1.15739  \pm  0.00050 $\\
 3000  & $18.230811  \pm  0.000093 $  & $19.498275  \pm  0.000058   $& $ 1.15748  \pm  0.00052 $\\
 3500  & $18.260909  \pm  0.000081 $  & $19.521898  \pm  0.000053   $& $ 1.15753  \pm  0.00058 $\\
 4000  & $18.287471  \pm  0.000069 $  & $19.543763  \pm  0.000049   $& $ 1.15772  \pm  0.00066 $\\
 5000  & $18.331323  \pm  0.000049 $  & $19.582986  \pm  0.000043   $& $ 1.15749  \pm  0.00083 $\\
 6000  & $18.364968  \pm  0.000034 $  & $19.617183  \pm  0.000037   $& $ 
                1.1566\hphantom{0}  \pm  0.0010\hphantom{0} $\\
 7000  & $18.389939  \pm  0.000021 $  & $19.647298  \pm  0.000032   $& $ 
                1.1586\hphantom{0}  \pm  0.0013\hphantom{0} $\\
 8000  & $18.407212  \pm  0.000011 $  & $19.673754  \pm  0.000027   $& $ 
                1.1582\hphantom{0}  \pm  0.0017\hphantom{0} $\\
\hline
\end{tabular}
\end{center}
\caption{Raw data for $N_{tot} = 20000$ and $N_{tot} = 40000$ and combined
estimate of $\gamma$ for various values of $N_{min}$.}
\label{tabella_gamma2}
 \end{table}

\end{document}